\documentclass{svproc}
\usepackage{url}
\usepackage{cite}
\usepackage{multirow}
\usepackage[pdftex]{graphicx}
\usepackage[export]{adjustbox}
\usepackage{hyperref}

\usepackage{booktabs,subcaption,amsfonts,dcolumn}

\usepackage{fixltx2e}
\usepackage{subcaption}
\begin{document}
\mainmatter              
\title{A BERT-Based Transfer Learning Approach for Hate Speech Detection in Online Social Media}
\titlerunning{A BERT-based hate speech detection approach in social media}  
%
\author{Marzieh Mozafari\inst{1} \and Reza Farahbakhsh\inst{1} \and No\"{e}l  Crespi\inst{1}}
\authorrunning{Marzieh Mozafari et al.} 
%
\tocauthor{Marzieh Mozafari, Reza Farahbakhsh,  No\"{e}l Crespi}
\institute{CNRS UMR5157, T\'el\'ecom SudParis, Institut Polytechnique de Paris, \'Evry, France \\
\email{\{marzieh.mozafari, reza.farahbakhsh, noel.crespi\}@telecom-sudparis.eu}}

\maketitle              

\begin{abstract}
Generated hateful and toxic content by a portion of users in social media is a rising phenomenon that motivated researchers to dedicate substantial efforts to the challenging direction of hateful content identification. We not only need an efficient automatic hate speech detection model based on advanced machine learning and natural language processing, but also a sufficiently large amount of annotated data to train a model. The lack of a sufficient amount of labelled hate speech data, along with the existing biases, has been the main issue in this domain of research. To address these needs, in this study we introduce a novel transfer learning approach based on an existing pre-trained language model called BERT (Bidirectional Encoder Representations from Transformers). More specifically, we investigate the ability of BERT at capturing hateful context within social media content by using new fine-tuning methods based on transfer learning. To evaluate our proposed approach, we use two publicly available datasets that have been annotated for racism, sexism, hate, or offensive content on Twitter. The results show that our solution obtains considerable performance on these datasets in terms of precision and recall in comparison to existing approaches. Consequently, our model can capture some biases in data annotation and collection process and can potentially lead us to a more accurate model.
	
\keywords{hate speech detection, transfer learning, language modeling, BERT, fine-tuning, NLP, social media.}
\end{abstract}

\section{Introduction}
\label{sec:Introduction}
\vspace{-1.5ex}

People are increasingly using social networking platforms such as Twitter, Facebook, YouTube, etc. to communicate their opinions and share information. Although the interactions among users on these platforms can lead to constructive conversations, they have been increasingly exploited for the propagation of abusive language and the organization of hate-based activities \cite{BadjatiyaG0V17, burnap2015}, especially due to the mobility and anonymous environment of these online platforms. Violence attributed to online hate speech has increased worldwide. For example, in the UK, there has been a significant increase in hate speech towards the immigrant and Muslim communities following the UK's leaving the EU and the Manchester and London attacks\footnote{\scriptsize{Anti-muslim hate crime surges after Manchester and London Bridge attacks (2017): \url{https://www.theguardian.com}}}.
The US also has been a marked increase in hate speech and related crime following the Trump election\footnote{\scriptsize{A.: Hate on the rise after Trump’s election: \url{http://www.newyorker.com}}}. Therefore, governments and social network platforms confronting the trend must have tools to detect aggressive behavior in general, and hate speech in particular, as these forms of online aggression not only poison the social climate of the online communities that experience it, but can also provoke physical violence and serious harm \cite{burnap2015}.

Recently, the problem of online abusive detection has attracted scientific attention. Proof of this is the creation of the third Workshop on Abusive Language Online\footnote{\scriptsize \url{https://sites.google.com/view/alw3/home}} or Kaggle’s Toxic Comment Classification Challenge that gathered 4,551 teams\footnote{\scriptsize \url{https://www.kaggle.com/c/jigsaw-toxic-comment-classification-challenge/}} in 2018 to detect different types of toxicities (threats, obscenity, etc.). In the scope of this work, we mainly focus on the term hate speech as abusive content in social media, since it can be considered a broad umbrella term for numerous kinds of insulting user-generated content. Hate speech is commonly defined as any communication criticizing a person or a group based on some characteristics such as gender, sexual orientation, nationality, religion, race, etc. Hate speech detection is not a stable or simple target because misclassification of regular conversation as hate speech can severely affect users’ freedom of expression and reputation, while misclassification of hateful conversations as unproblematic would maintain the status of online communities as unsafe environments \cite{DavidsonBhattacharya2019}.

To detect online hate speech, a large number of scientific studies have been dedicated by using Natural Language Processing (NLP) in combination with Machine Learning (ML) and Deep Learning (DL) methods \cite{Nobata2016, mehdad2016, waseemhovy2016, gamback2017, Zhang2018, BadjatiyaG0V17}. Although supervised machine learning-based approaches have used different text mining-based features such as surface features, sentiment analysis, lexical resources, linguistic features, knowledge-based features or user-based and platform-based metadata \cite{Fortuna2018, Davidson2017, Waseem2018}, they necessitate a well-defined feature extraction approach. The trend now seems to be changing direction, with deep learning models being used for both feature extraction and the training of classifiers. These newer models are applying deep learning approaches such as Convolutional Neural Networks (CNNs), Long Short-Term Memory Networks (LSTMs), etc.\cite{gamback2017, BadjatiyaG0V17} to enhance the performance of hate speech detection models, however, they still suffer from lack of labelled data or inability to improve generalization property.

 Here, we propose a transfer learning approach for hate speech understanding using a combination of the unsupervised pre-trained model  BERT \cite{bert2019} and some new supervised fine-tuning strategies. As far as we know, it is the first time that such exhaustive fine-tuning strategies are proposed along with a generative pre-trained language model to transfer learning to low-resource hate speech languages and improve performance of the task.  In summary:
\begin{itemize}
  	\item[$\bullet$] We propose a transfer learning approach using the pre-trained language model BERT learned on English Wikipedia and BookCorpus to enhance hate speech detection on publicly available benchmark datasets. Toward that end, for the first time, we introduce new fine-tuning strategies to examine the effect of different embedding layers of BERT in hate speech detection.
  	\item[$\bullet$] Our experiment results show that using the pre-trained BERT model and fine-tuning it on the downstream task by leveraging syntactical and contextual information of all BERT's transformers outperforms previous works in terms of precision, recall, and F1-score. Furthermore, examining the results shows the ability of our model to detect some biases in the process of collecting or annotating datasets. It can be a valuable clue in using pre-trained BERT model for debiasing hate speech datasets in future studies.
\end{itemize}

\section{Previous Works}
\label{sec:Liturature}
\vspace{-1.5ex}
Here, the existing body of knowledge on online hate speech and offensive language and transfer learning is presented.

\noindent\textbf{Online Hate Speech and Offensive Language:}
Researchers have been studying hate speech on social media platforms such as Twitter \cite{Davidson2017}, Reddit \cite{Olteanu2018, Mittos2019}, and YouTube \cite{Ottoni2018} in the past few years. The features used in traditional machine learning approaches are the main aspects distinguishing different methods, and surface-level features such as bag of words, word-level and character-level $n$-grams, etc. have proven to be the most predictive features \cite{Nobata2016, mehdad2016, waseemhovy2016}. Apart from features, different algorithms such as Support Vector Machines \cite{Malmasi2018}, Naive Baye \cite{burnap2015}, and Logistic Regression \cite{waseemhovy2016, Davidson2017}, etc. have been applied for classification purposes. Waseem et al. \cite{waseemhovy2016} provided a test with a list of criteria based on the work in Gender Studies and Critical Race Theory (CRT) that can annotate a corpus of more than $16k$ tweets as racism, sexism, or neither. To classify tweets, they used a logistic regression model with different sets of features, such as word and character $n$-grams up to 4, gender, length, and location. They found that their best model produces character $n$-gram as the most indicative features, and using location or length is detrimental. Davidson et al. \cite{Davidson2017} collected a $24K$ corpus of tweets containing hate speech keywords and labelled the corpus as hate speech, offensive language, or neither by using crowd-sourcing and extracted different features such as $n$-grams, some tweet-level metadata such as the number of hashtags, mentions, retweets, and URLs, Part Of Speech (POS) tagging, etc. Their experiments on different multi-class classifiers showed that the Logistic Regression with L2 regularization performs the best at this task. Malmasi et al. \cite{Malmasi2018} proposed an ensemble-based system that uses some linear SVM classifiers in parallel to distinguish hate speech from general profanity in social media. 

As one of the first attempts in neural network models, Djuric et al. \cite{Djuric2015} proposed a two-step method including a continuous bag of words model to extract paragraph2vec embeddings and a binary classifier trained along with the embeddings to distinguish between hate speech and clean content. Badjatiya et al. \cite{BadjatiyaG0V17} investigated three deep learning architectures, FastText, CNN, and LSTM, in which they initialized the word embeddings with either random or GloVe embeddings. Gambäck et al. \cite{gamback2017} proposed a hate speech classifier based on CNN model trained on different feature embeddings such as word embeddings and character $n$-grams. Zhang et al. \cite{Zhang2018} used a CNN+GRU (Gated Recurrent Unit network) neural network model initialized with pre-trained word2vec embeddings to capture both word/character combinations (e. g., $n$-grams, phrases) and word/character dependencies (order information). Waseem et al. \cite{Waseem2018} brought a new insight to hate speech and abusive language detection tasks by proposing a multi-task learning framework to deal with datasets across different annotation schemes, labels, or geographic and cultural influences from data sampling. Founta et al. \cite{Founta2019} built a unified classification model that can efficiently handle different types of abusive language such as cyberbullying, hate, sarcasm, etc. using raw text and domain-specific metadata from Twitter. Furthermore, researchers have recently focused on the bias derived from the hate speech training datasets \cite{WaseemDavidson2017, DavidsonBhattacharya2019, wiegand2019}. Davidson et al. \cite{DavidsonBhattacharya2019} showed that there were systematic and substantial racial biases in five benchmark Twitter datasets annotated for offensive language detection. Wiegand et al. \cite{wiegand2019} also found that classifiers trained on datasets containing more implicit abuse (tweets with some abusive words) are more affected by biases rather than once trained on datasets with a high proportion of explicit abuse samples (tweets containing sarcasm, jokes, etc.). 
\noindent\textbf{Transfer Learning:}
Pre-trained vector representations of words, embeddings, extracted from vast amounts of text data have been encountered in almost every language-based tasks with promising results. Two of the most frequently used context-independent neural embeddings are word2vec and Glove extracted from shallow neural networks. The year 2018 has been an inflection point for different NLP tasks thanks to remarkable breakthroughs: Universal Language Model Fine-Tuning (ULMFiT) \cite{Ruder2018}, Embedding from Language Models (ELMO) \cite{Matthew_2018}, OpenAI’ s Generative Pre-trained Transformer (GPT) \cite{Radford2018}, and Google’s BERT model \cite{bert2019}. Howard et al. \cite{Ruder2018} proposed ULMFiT which can be applied to any NLP task by pre-training a universal language model on a general-domain corpus and then fine-tuning the model on target task data using discriminative fine-tuning. Peters et al. \cite{Matthew_2018} used a bi-directional LSTM trained on a specific task to present context-sensitive representations of words in word embeddings by looking at the entire sentence. Radford et al. \cite{Radford2018} and Devlin et al. \cite{bert2019} generated two transformer-based language models, OpenAI GPT and BERT respectively. OpenAI GPT \cite{Radford2018} is an unidirectional language model while BERT \cite{bert2019} is the first deeply bidirectional, unsupervised language representation, pre-trained using only a plain text corpus. BERT has two novel prediction tasks: Masked LM and Next Sentence Prediction. The pre-trained BERT model significantly outperformed ELMo and OpenAI GPT in a series of downstream tasks in NLP \cite{bert2019}. Identifying hate speech and offensive language is a complicated task due to the lack of undisputed labelled data \cite{Malmasi2018} and the inability of surface features to capture the subtle semantics in text. To address this issue, we use the pre-trained language model BERT for hate speech classification and try to fine-tune specific task by leveraging information from different transformer encoders.

\section{Methodology}
\vspace{-1.5ex}
Here, we analyze the BERT transformer model on the hate speech detection task.  BERT is a multi-layer bidirectional transformer encoder trained on the English Wikipedia and the Book Corpus containing 2,500M and 800M tokens, respectively, and has two models named BERT\textsubscript{base} and BERT\textsubscript{large}. BERT\textsubscript{base} contains an encoder with 12 layers (transformer blocks), 12 self-attention heads, and 110 million parameters whereas BERT\textsubscript{large} has 24 layers, 16 attention heads, and 340 million parameters. Extracted embeddings from BERT\textsubscript{base} have 768 hidden dimensions \cite{bert2019}. As the BERT model is pre-trained on general corpora, and for our hate speech detection task we are dealing with social media content, therefore as a crucial step, we have to analyze the contextual information extracted from BERT' s pre-trained layers and then fine-tune it using annotated datasets. By fine-tuning we update weights using a labelled dataset that is new to an already trained model. As an input and output, BERT takes a sequence of tokens in maximum length 512 and produces a representation of the sequence in a 768-dimensional vector. BERT inserts at most two segments to each input sequence, [CLS] and [SEP]. [CLS] embedding is the first token of the input sequence and contains the special classification embedding which we take the first token [CLS] in the final hidden layer as the representation of the whole sequence in hate speech classification task. The [SEP] separates segments and we will not use it in our classification task. To perform the hate speech detection task, we use BERT\textsubscript{base} model to classify each tweet as Racism, Sexism, Neither or Hate, Offensive, Neither in our datasets. In order to do that, we focus on fine-tuning the pre-trained BERT\textsubscript{base} parameters. By fine-tuning, we mean training a classifier with different layers of 768 dimensions on top of the pre-trained BERT\textsubscript{base} transformer to minimize task-specific parameters.

\subsection{Fine-Tuning Strategies}
\begin{figure}[t]
\vspace{-6ex}
\hspace{-6ex}
\scalebox{0.8}{%
\begin{subfigure}[b]{0.4\textwidth}
  \includegraphics[width=0.75\linewidth]{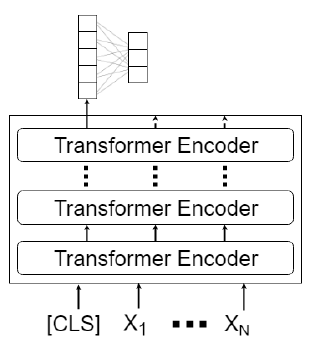}  
    \caption{\footnotesize{BERT\textsubscript{base} fine-tuning}} 
    \label{fig:1} 
\end{subfigure}\hspace{-0.9em}%
\hspace{-6ex}
\begin{subfigure}[b]{.4\textwidth}
  \includegraphics[width=0.75\linewidth]{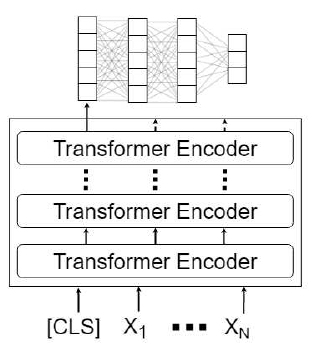}  
    \caption{\footnotesize{Insert nonlinear layers}} 
    \label{fig:2} 
\end{subfigure}\hspace{-0.9em}%
\hspace{-6ex}
\begin{subfigure}[b]{.4\textwidth}
  \includegraphics[width=0.75\linewidth]{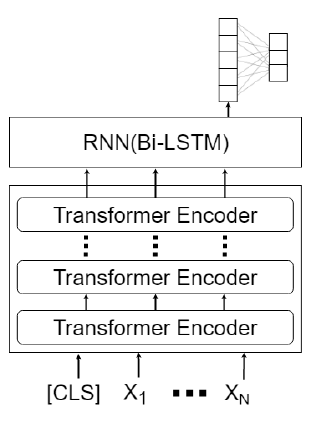}  
    \caption{\footnotesize{Insert Bi-LSTM layer}} 
    \label{fig:3} 
\end{subfigure}\hspace{-0.7em}%
\hspace{-6ex}
\begin{subfigure}[b]{.4\textwidth}
  \includegraphics[width=1.1\linewidth]{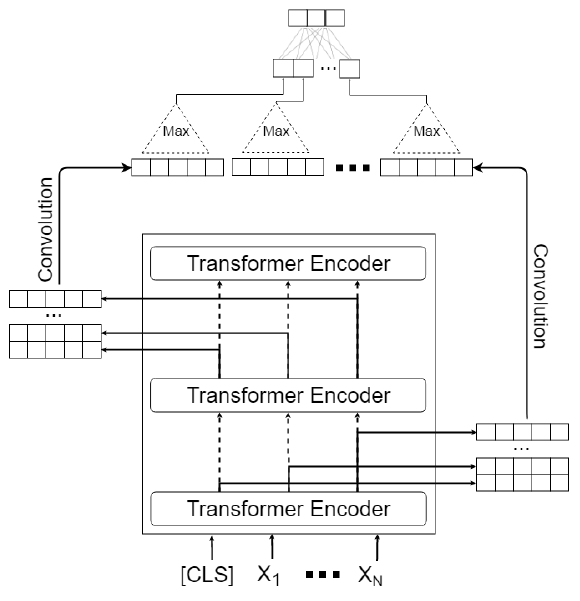}  
    \caption{\footnotesize{Insert CNN layer}} 
    \label{fig:4} 
\end{subfigure}
}
  \caption{\footnotesize{Fine-tuning strategies}}
  \label{fig:finetuning_strategies} 
\vspace{-5ex}
\end{figure}

Different layers of a neural network can capture different levels of syntactic and semantic information. The lower layer of the BERT model may contain more general information whereas the higher layers contain task-specific information \cite{bert2019}, and we can fine-tune them with different learning rates. Here, four different fine-tuning approaches are implemented that exploit pre-trained BERT\textsubscript{base} transformer encoders for our classification task. More information about these transformer encoders' architectures are presented in \cite{bert2019}. In the fine-tuning phase, the model is initialized with the pre-trained parameters and then are fine-tuned using the labelled datasets. Different fine-tuning approaches on the hate speech detection task are depicted in Figure \ref{fig:finetuning_strategies}, in which $X_{i}$ is the vector representation of token $i$ in a tweet sample, and are explained in more detail as follows:

\textbf{1. BERT based fine-tuning:}
In the first approach, which is shown in Figure \ref{fig:1}, very few changes are applied to the BERT\textsubscript{base}. In this architecture, only the [CLS] token output provided by BERT is used. The [CLS] output, which is equivalent to the [CLS] token output of the 12th transformer encoder, a vector of size 768, is given as input to a fully connected network without hidden layer. The softmax activation function is applied to the hidden layer to classify.

\textbf{2. Insert nonlinear layers:}
Here, the first architecture is upgraded and an architecture with a more robust classifier is provided in which instead of using a fully connected network without hidden layer, a fully connected network with two hidden layers in size 768  is used. The first two layers use the Leaky Relu activation function with negative slope = 0.01, but the final layer, as the first architecture, uses softmax activation function as shown in Figure \ref{fig:2}.

\textbf{3. Insert Bi-LSTM layer:}
Unlike previous architectures that only use [CLS] as the input for the classifier, in this architecture all outputs of the latest transformer encoder are used in such a way that they are given as inputs to a bidirectional recurrent neural network (Bi-LSTM) as shown in Figure \ref{fig:3}. After processing the input, the network sends the final hidden state to a fully connected network that performs classification using the softmax activation function.

\textbf{4. Insert CNN layer:}
In this architecture shown in Figure \ref{fig:4}, the outputs of all transformer encoders are used instead of using the output of the latest transformer encoder. So that the output vectors of each transformer encoder are concatenated, and a matrix is produced. The convolutional operation is performed with a window of size (3, hidden size of BERT which is 768 in BERT\textsubscript{base} model) and the maximum value is generated for each transformer encoder by applying max pooling on the convolution output. By concatenating these values, a vector is generated which is given as input to a fully connected network. By applying softmax on the input, the classification operation is performed.

\section{Experiments and Results}
\vspace{-1.5ex}
We first introduce datasets used in our study and then investigate the different fine-tuning strategies for hate speech detection task. We also include the details of our implementation and error analysis in the respective subsections.

\subsection{Dataset Description}
We evaluate our method on two widely-studied datasets provided by Waseem and Hovey \cite{waseemhovy2016} and Davidson et al. \cite{Davidson2017}. Waseem and Hovy \cite{waseemhovy2016} collected $16k$ of tweets based on an initial ad-hoc approach that searched common slurs and terms related to religious, sexual, gender, and ethnic minorities. They annotated their dataset manually as racism, sexism, or neither. To extend this dataset, Waseem \cite{waseem2016} also provided another dataset containing $6.9k$ of tweets annotated with both expert and crowdsourcing users as racism, sexism, neither, or both. Since both datasets are overlapped partially and they used the same strategy in definition of hateful content, we merged these two datasets following Waseem et al. \cite{Waseem2018} to make our imbalance data a bit larger. Davidson et al. \cite{Davidson2017} used the Twitter API to accumulate 84.4 million tweets from 33,458 twitter users containing particular terms from a pre-defined lexicon of hate speech words and phrases, called Hatebased.org. To annotate collected tweets as Hate, Offensive, or Neither, they randomly sampled $25k$ tweets and asked users of CrowdFlower crowdsourcing platform to label them. In detail, the distribution of different classes in both datasets will be provided in Subsection \ref{implementation}.

\subsection{Pre-Processing}
We find mentions of users, numbers, hashtags, URLs and common emoticons and replace them with the tokens \textless user\textgreater,\textless number\textgreater,\textless hashtag\textgreater,\textless url\textgreater,\textless emoticon\textgreater. We also find elongated words and convert them into short and standard format; for example, converting yeeeessss to yes. With hashtags that include some tokens without any with space between them, we replace them by their textual counterparts; for example, we convert hashtag ``\#notsexist" to ``not sexist". All punctuation marks, unknown uni-codes and extra delimiting characters are removed, but we keep all stop words because our model trains the sequence of words in a text directly. We also convert all tweets to lower case. 

\subsection{Implementation and Results Analysis }
\label{implementation}
For the implementation of our neural network, we used pytorch-pretrained-bert library
containing the pre-trained BERT model, text tokenizer, and pre-trained WordPiece. As the implementation environment, we use Google Colaboratory tool
which is a free research tool with a Tesla K80 GPU and 12G RAM. Based on our experiments, we trained our classifier with a batch size of 32 for 3 epochs. The dropout probability is set to 0.1 for all layers. Adam optimizer is used with a learning rate of 2e-5. As an input, we tokenized each tweet with the BERT tokenizer. It contains invalid characters removal, punctuation splitting, and lowercasing the words. Based on the original BERT \cite{bert2019}, we split words to subword units using WordPiece tokenization. As tweets are short texts, we set the maximum sequence length to 64 and in any shorter or longer length case it will be padded with zero values or truncated to the maximum length.

We consider 80\% of each dataset as training data to update the weights in the fine-tuning phase, 10\% as validation data to measure the out-of-sample performance of the model during training, and 10\% as test data to measure the out-of-sample performance after training. To prevent overfitting, we use stratified sampling to select 0.8, 0.1, and 0.1 portions of tweets from each class (racism/sexism/neither or hate/offensive/neither) for train, validation, and test. Classes' distribution of train, validation, and test datasets are shown in Table \ref{tab:table1}.

\begin{table}
\vspace{-5ex}
\caption{\footnotesize Dataset statistics of the both Waseem-dataset (\subref{class_distribution_waseem}) and Davidson-dataset (\subref{class_distribution_davidson}). Splits are produced using stratified sampling to select 0.8, 0.1, and 0.1 portions of tweets from each class (racism/sexism/neither or hate/offensive/neither) for train, validation, and test samples, respectively.}
\label{tab:table1}
\begin{subtable}{0.4\textwidth}
\scriptsize
\begin{tabular}{lcccc}
\hline
                    & \textbf{Racism} & \textbf{Sexism} & \textbf{Neither} & \textbf{Total} \\ \hline
\textbf{Train}      & 1693            & 3337            & 10787            & 15817          \\
\textbf{Validation} & 210             & 415             & 1315             & 1940           \\
\textbf{Test}       & 210             & 415             & 1315             & 1940           \\ \hline
\textbf{Total}      & 2113            & 4167            & 13417            &                \\ \hline
\end{tabular}
\caption{\footnotesize Waseem-dataset.}
\label{class_distribution_waseem}
\end{subtable}
\hspace{\fill}
\begin{subtable}{0.4\textwidth}
\flushright
\scriptsize
\begin{tabular}{lcccc}
\hline
                    & \textbf{Hate} & \textbf{Offensive} & \textbf{Neither} & \textbf{Total} \\ \hline
\textbf{Train}      & 1146          & 15354              & 3333             & 19832          \\
\textbf{Validation} & 142           & 1918               & 415              & 2475           \\
\textbf{Test}       & 142           & 1918               & 415              & 2475           \\ \hline
\textbf{Total}      & 1430          & 19190              & 4163             &                \\ \hline
\end{tabular}
\caption{\footnotesize Davidson-dataset.}
\label{class_distribution_davidson}
\end{subtable}
\vspace{-5ex}
\end{table}

As it is understandable from Tables \ref{tab:table1}(\subref{class_distribution_waseem}) and \ref{tab:table1}(\subref{class_distribution_davidson}), we are dealing with imbalance datasets with various classes’ distribution. Since hate speech and offensive languages are real phenomena, we did not perform oversampling or undersampling techniques to adjust the classes’ distribution and tried to supply the datasets as realistic as possible. We evaluate the effect of different fine-tuning strategies on the performance of our model. Table \ref{bert-fine-tune} summarized the obtained results for fine-tuning strategies along with the official baselines. We use Waseem and Hovy \cite{waseemhovy2016}, Davidson et al. \cite{Davidson2017}, and Waseem et al. \cite{Waseem2018} as baselines and compare the results with our different fine-tuning strategies using pre-trained BERT\textsubscript{base} model. The evaluation results are reported on the test dataset and on three different metrics: precision, recall, and weighted-average F1-score. We consider weighted-average F1-score as the most robust metric versus class imbalance, which gives insight into the performance of our proposed models. According to Table \ref{bert-fine-tune}, F1-scores of all BERT based fine-tuning strategies except BERT +  nonlinear classifier on top of BERT are higher than the baselines. Using the pre-trained BERT model as initial embeddings and fine-tuning the model with a fully connected linear classifier (BERT\textsubscript{base}) outperforms previous baselines yielding F1-score of 81\% and 91\% for datasets of Waseem and Davidson respectively.  Inserting a CNN to pre-trained BERT model for fine-tuning on downstream task provides the best results as F1- score of 88\% and 92\% for datasets of Waseem and Davidson and it clearly exceeds the baselines. Intuitively, this makes sense that combining all pre-trained BERT layers with a CNN yields better results in which our model uses all the information included in different layers of pre-trained BERT during the fine-tuning phase. This information contains both syntactical and contextual features coming from lower layers to higher layers of BERT.

\begin{table}[!h]
\vspace{-4ex}
\footnotesize
\center
\caption{\footnotesize{Results on the trial data using pre-trained BERT model with different fine-tuning strategies and comparison with results in the literature.}}
\label{bert-fine-tune}
\scalebox{1}{%
\begin{tabular}{lllll}
\hline
\textbf{Method} & \textbf{Datasets} & \textbf{Precision(\%)} & \textbf{Recall(\%)} & \textbf{F1-Score(\%)} \\ \hline
Waseem and Hovy\cite{waseemhovy2016} & Waseem & 72.87 & 77.75 & 73.89 \\ \hline
Davidson et al.\cite{Davidson2017} & Davidson & 91 & 90 & 90 \\ \hline
\multirow{2}{*}{Waseem et al.\cite{Waseem2018}} & Waseem &  - &  - & 80 \\ \cline{2-5} 
 & Davidson & - & - & 89 \\ \hline
\multirow{2}{*}{BERT\textsubscript{base}} & Waseem & 81 & 81 & 81 \\ \cline{2-5} 
 & Davidson & 91 & 91 & 91 \\ \hline
\multirow{2}{*}{BERT\textsubscript{base} + Nonlinear Layers} & Waseem & 73 & 85 & 76  \\ \cline{2-5} 
 & Davidson & 76 &78  & 77 \\ \hline
 \multirow{2}{*}{BERT\textsubscript{base} + LSTM} & Waseem &87 & 86 & 86 \\ \cline{2-5} 
 & Davidson & 91  &92  & 92\\ \hline
\multirow{2}{*}{BERT\textsubscript{base} + CNN} & Waseem & \textbf{89}  & \textbf{87}  & \textbf{88}  \\ \cline{2-5} 
 & Davidson & \textbf{92}  & \textbf{92} &  \textbf{92} \\ \hline
\end{tabular}%
}
\vspace{-4ex}
\end{table}

\subsection{Error Analysis}
\begin{figure}[h]
\vspace{-6ex}
  \hspace{-0.6cm}
  \centering
  \begin{minipage}[b]{0.45\linewidth}
    \includegraphics[width=1.2\linewidth]{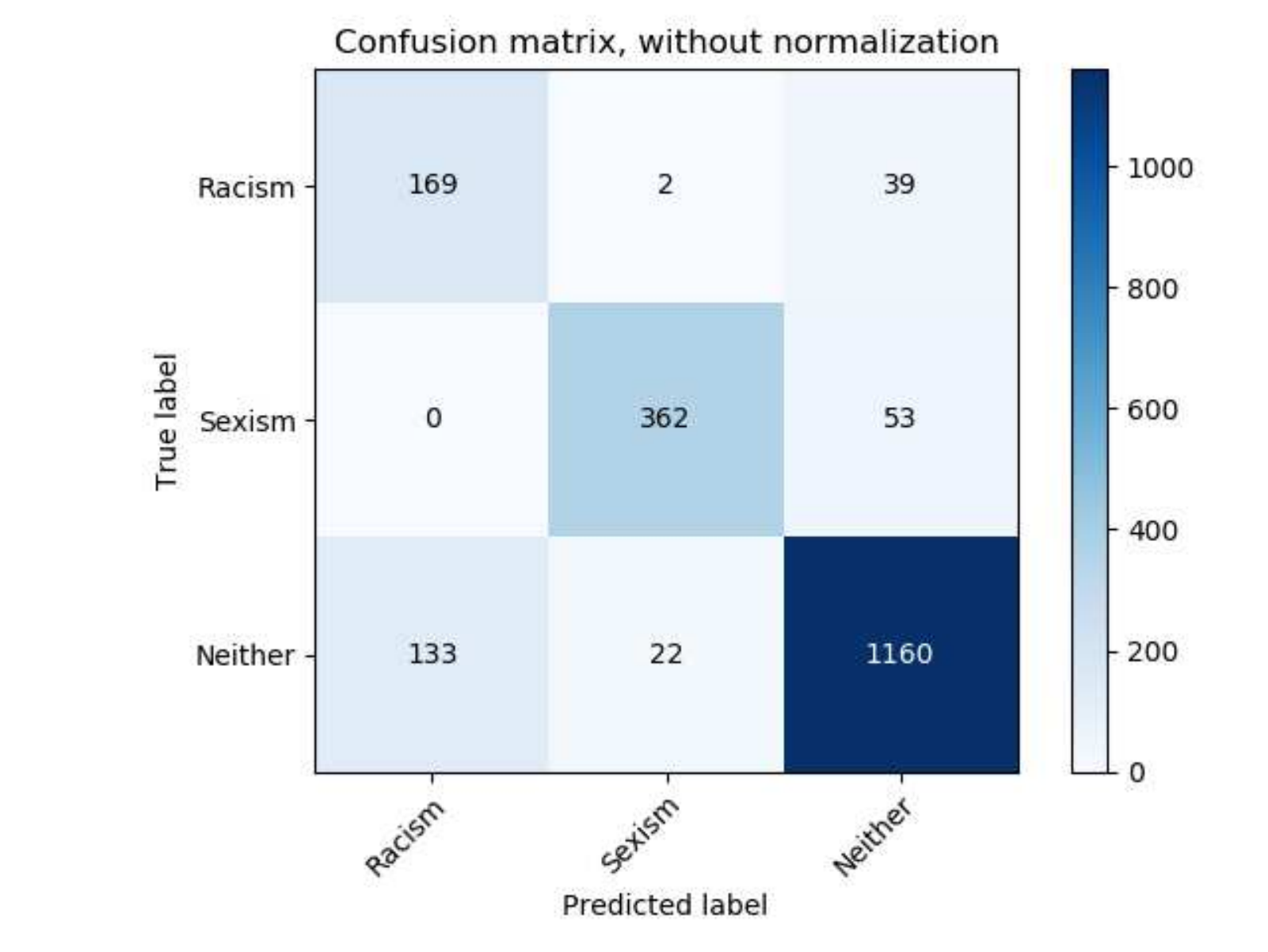}
    \caption{\scriptsize{ Waseem-datase's confusion matrix}}
    \label{fig:Zaraak}
  \end{minipage}
  \hfill
  \hspace{-1cm}
  \begin{minipage}[b]{0.45\linewidth}
    \includegraphics[width=1.2\linewidth]{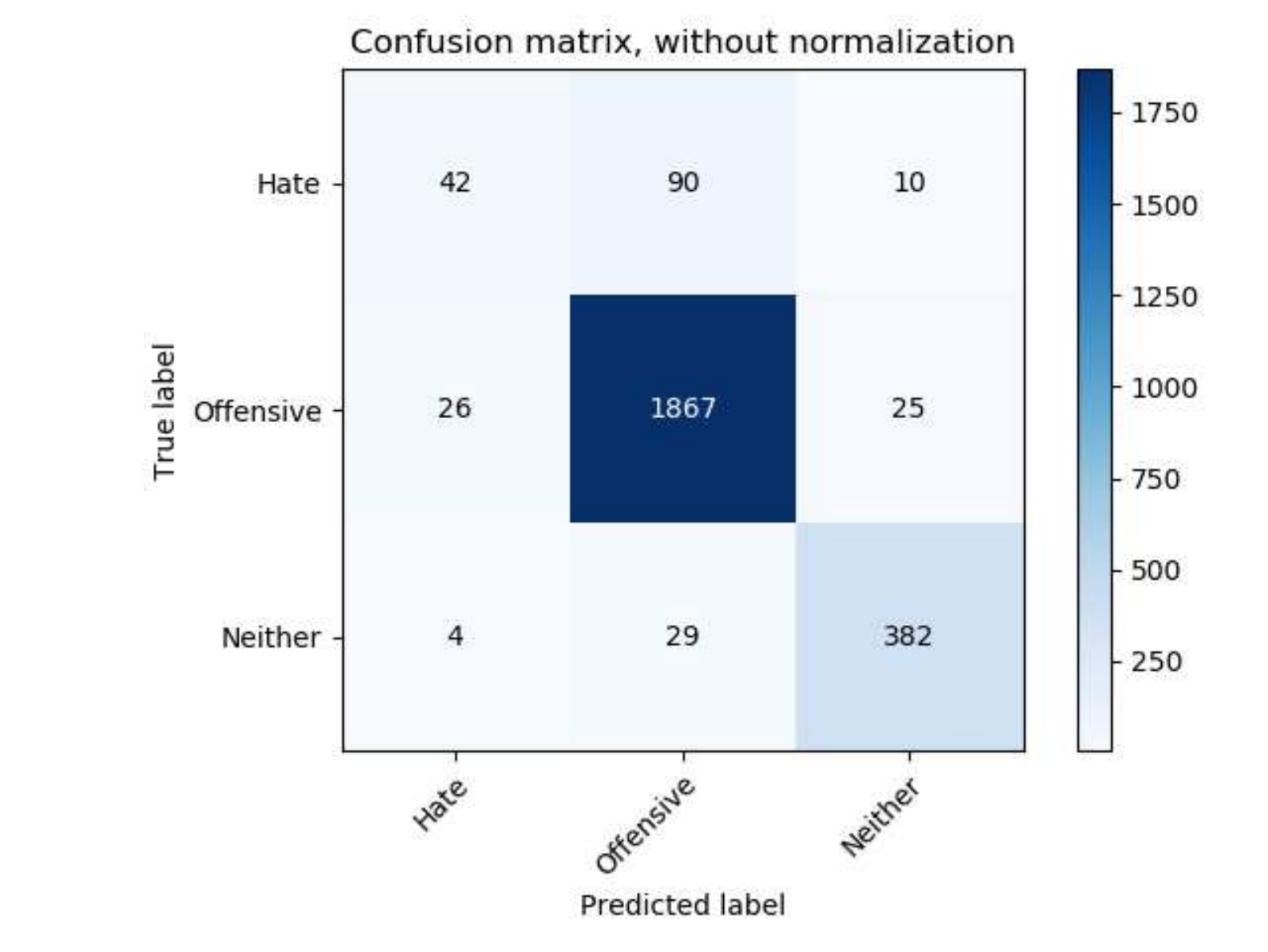}
    \caption{\scriptsize{ Davidson-dataset's confusion matrix}}
    \label{fig:Davidson}
  \end{minipage}
\vspace{-5ex}
\end{figure}

 Although we have very interesting results in term of recall, the precision of the model shows the portion of false detection we have. To understand better this phenomenon, in this section we perform a deep analysis on the error of the model. We investigate the test datasets and their confusion matrices resulted from the BERT\textsubscript{base} + CNN model as the best fine-tuning approach; depicted in Figures \ref{fig:Zaraak} and \ref{fig:Davidson}. According to Figure \ref{fig:Zaraak} for Waseem-dataset, it is obvious that the model can separate sexism from racism content properly. Only two samples belonging to racism class are misclassified as sexism and none of the sexism samples are misclassified as racism. A large majority of the errors come from misclassifying hateful categories (racism and sexism) as hatless (neither) and vice versa. 0.9\% and 18.5\% of all racism samples are misclassified as sexism and neither respectively whereas it is 0\% and 12.7\% for sexism samples. Almost 12\% of neither samples are misclassified as racism or sexism. As Figure  \ref{fig:Davidson} makes clear for Davidson-dataset, the majority of errors are related to hate class where the model misclassified hate content as offensive in 63\% of the cases. However, 2.6\% and 7.9\% of offensive and neither samples are misclassified respectively.

To understand better the mislabeled items by our model, we did a manual inspection on a subset of the data and record some of them in Tables \ref{waseem_error} and \ref{davidson_error}. Considering the words such as ``daughters", ``women", and ``burka" in tweets with IDs 1 and 2 in Table \ref{waseem_error}, it can be understood that our BERT based classifier is confused with the contextual semantic between these words in the samples and misclassified them as sexism because they are mainly associated to femininity. In some cases containing implicit abuse (like subtle insults) such as tweets with IDs 5 and 7,  our model cannot capture the hateful/offensive content and therefore misclassifies. It should be noticed that even for a human it is difficult to discriminate against this kind of implicit abuses.

\begin{table}[!th]
\vspace{-4ex}
\footnotesize
\center
\caption{\footnotesize{Misclassified samples from Waseem-dataset.}}
\label{waseem_error}
\scalebox{0.83}{%
\begin{tabular}{clcc}
\hline
\textbf{ID} & \multicolumn{1}{c}{\textbf{Tweet}} & \textbf{Annotated} & \textbf{Predicted} \\ \hline
1 & @user Good tweet. But they actually start selling their daughters at 9. & Racism & Sexism \\ \hline
2 & \begin{tabular}[c]{@{}l@{}}RT @user: Are we going to continue seeing the oppression of women or are we\\ going to make a stand? \#BanTheBurka http://t.co/hZDx8mlvTv.\end{tabular} & Racism & Sexism \\ \hline
3 & \begin{tabular}[c]{@{}l@{}}RT @user: @user my comment was sexist, but I'm not personally, always a sexist.
\end{tabular} & Sexism & Neither \\ \hline
4 & \begin{tabular}[c]{@{}l@{}}RT @user: @user Ah, you're a \#feminist? Seeing \#sexism everywhere then, do\\check my tweets before you call me \#sexist \end{tabular} & Sexism & Neither \\ \hline
\multicolumn{1}{c}{5} & @user By hating the ideology that enables it, that is what I'm doing. & Racism & Neither \\ \hline
\end{tabular}%
}
\vspace{-4ex}
\end{table}

\begin{table}[!th]
\footnotesize
\center
\caption{\footnotesize{Misclassified samples from Davidson-dataset.}}
\label{davidson_error}
\scalebox{0.85}{%
\begin{tabular}{clcc}
\hline
\textbf{ID} & \multicolumn{1}{c}{\textbf{Tweet}} & \textbf{Annotated} & \textbf{Predicted} \\ \hline
6 & \begin{tabular}[c]{@{}l@{}}@user: If you claim Macklemore is your favorite rapper I'm also assuming you \\ watch the WNBA on your free time faggot\end{tabular} & Hate & Offensive \\ \hline
7 & \begin{tabular}[c]{@{}l@{}}@user: Some black guy at my school asked if there were colored printers in the\\ library. "It's 2014 man you can use any printer you want” I said.\end{tabular} & Hate & Neither \\ \hline
8 & RT @user: @user typical coon activity. & Hate & Neither \\ \hline
9 & \begin{tabular}[c]{@{}l@{}}@user: @user @user White people need those weapons to defend themselves\\from the subhuman trash your sort unleashes on us.\end{tabular} & Neither & Hate \\ \hline
10 & \begin{tabular}[c]{@{}l@{}}RT @user: Finally! Warner Bros. making superhero films starring a woman,\\person of color and actor who identifies as ""queer"";\end{tabular} & Neither & Offensive \\ \hline
\end{tabular}%
}
\vspace{-6ex}
\end{table}

By examining more samples and with respect to recently studies \cite{DavidsonBhattacharya2019, sap2019, wiegand2019}, it is clear that many errors are due to biases from data collection \cite{wiegand2019} and rules of annotation \cite{sap2019} and not the classifier itself. Since Waseem et al.\cite{waseemhovy2016} created a small ad-hoc set of keywords and Davidson et al.\cite{Davidson2017} used a large crowdsourced dictionary of keywords (Hatebase lexicon) to sample tweets for training, they included some biases in the collected data. Especially for Davidson-dataset, some tweets with specific language (written within the African American Vernacular English) and geographic restriction (United States of America) are oversampled such as tweets containing disparage words ``nigga", ``faggot", ``coon", or ``queer", result in high rates of misclassification. However, these misclassifications do not confirm the low performance of our classifier because annotators tended to annotate many samples containing disrespectful words as hate or offensive without any presumption about the social context of tweeters such as the speaker’s identity or dialect, whereas they were just offensive or even neither tweets. Tweets IDs 6, 8, and 10
are some samples containing offensive words and slurs which arenot hate or offensive in all cases and writers of them used this type of language in their daily communications. Given these pieces of evidence, by considering the content of tweets, we can see in tweets IDs 3, 4, and 9 that our BERT-based classifier can discriminate tweets in which neither and implicit hatred content exist. One explanation of this observation may be the pre-trained general knowledge that exists in our model. Since the pre-trained BERT model is trained on general corpora, it has learned general knowledge from normal textual data without any purposely hateful or offensive language. Therefore, despite the bias in the data, our model can differentiate hate and offensive samples accurately by leveraging knowledge-aware language understanding that it has and it can be the main reason for high misclassifications of hate samples as offensive (in reality they are more similar to offensive rather than hate by considering social context, geolocation, and dialect of tweeters).

\section{Conclusion}
\vspace{-1.5ex}
Conflating hatred content with offensive or harmless language causes online automatic hate speech detection tools to flag user-generated content incorrectly. Not addressing this problem may bring about severe negative consequences for both platforms and users such as decreasement of platforms' reputation or users abandonment. Here, we propose a transfer learning approach advantaging the pre-trained language model BERT to enhance the performance of a hate speech detection system and to generalize it to new datasets. To that end, we introduce new fine-tuning strategies to examine the effect of different layers of BERT in hate speech detection task. The evaluation results indicate that our model outperforms previous works by profiting the syntactical and contextual information embedded in different transformer encoder layers of the BERT model using a CNN-based fine-tuning strategy. Furthermore, examining the results shows the ability of our model to detect some biases in the process of collecting or annotating datasets. It can be a valuable clue in using the pre-trained BERT model to alleviate bias in hate speech datasets in future studies, by investigating a mixture of contextual information embedded in the BERT’s layers and a set of features associated to the different type of biases in data.
\vspace{-1.5ex}
\bibliographystyle{spmpsci}
\bibliography{sample-bibliography}
\end{document}